\documentclass[twocolumn, eqsecnum, aps ]{revtex4}
\usepackage{graphicx}
\usepackage{dcolumn}

\begin{document}
\title{Application of Relativistic Coupled-cluster Theory to Heavy Atomic Systems with Strongly Interacting Configurations: {\it Hyperfine Interactions in $ ^{207}Pb^+$}} 
\vspace*{0.5cm}

\author{Bijaya K. Sahoo} 
\email{bijaya@iiap.res.in}
\author{Rajat K. Chaudhuri, B.P. Das\\
{\it Non-Accelerator Particle Physics Group\\Indian Institute of Astrophysics, Bangalore-34, India}\\
Holger Merlitz\\
{\it Forschungszentrum Karlsruhe GmbH, Institut f\"ur Nanotechnologie, \\
Postfach 3640, D-76021 Karlsruhe, Germany}\\
Debashis Mukherjee\\
{\it Department of Physical Chemistry \\Indian Association for Cultivation of Science, Kolkata - 700 032, India}}
\date{Received date; Accepted date}
\begin{abstract}

\noindent
This work presents a first time accurate calculation of the magnetic dipole
hyperfine structure constants for the ground state and some low-lying excited
states of Pb$^+$. By comparing different levels of approximation with
experimental results, we demonstrate the importance of correlation effects 
which reach beyond lower order relativistic many body perturbation theory. 
Employing relativistic coupled-cluster theory we obtain a quantitative understanding
of the core-polarization and correlation effects inherent in this system 
and observe completely different trends compared to $Ba^+$.   
\end{abstract} 
\maketitle

\noindent
Coupled-cluster theory has been used to study a wide range of many-body
systems\cite{bishop}. Although the non-relativistic version of this theory has
been very successfully applied to a variety of light atoms and
molecules\cite{kaldor01}, its extension to the relativistic regime is rather
recent\cite{kaldor02,merlitz}. There have been relatively few theoretical
studies of properties of heavy atomic systems based on the relativistic
coupled-cluster (RCC) theory.
Pb$^+$(Z=82) is the heaviest atomic
ion that has been trapped and cooled so far\cite{strumia,werth01}. The magnetic dipole hyperfine 
constants have been measured for the $6p ^2P_{1/2}$ and $6p ^2P_{3/2}$ states of this
ion\cite{werth02} and these data can be compared with 
 calculations of the corresponding quantities using RCC theory. Such
comparisons would indeed constitute an important test of this theory. 
The non-linear RCC in the singles and doubles approximation with partial
triples added in some cases has yielded results to an accuracy of about         
one percent for atoms and ions with a single s valence electron \cite{sahoo01,sahoo02,csur01}. However, the correlation effects in Pb$^+$ are expected to be much stronger as it has a 6p valence electron and two 6s electrons in its outermost core orbital.\\

\noindent
The hyperfine structure constant ({\it A})for the atomic state $|JM\rangle$ can be expressed in
terms of a reduced expectation value
\begin{eqnarray}
\label{eq:hfsc}
A &=& \mu_N [\frac {\mu_I}{I}] \frac {\langle J || T^{(1)} || J \rangle}{\sqrt{J
(J+1)(2J+1)}}
\end{eqnarray}
with $\mu_N$ being the nuclear magnetic moment and $[\frac {\mu_I}{I}]$ the
 Lande's nuclear g-factor ($g_I$).
 $T^{(1)}$ can be written as \cite{cheng}
\begin{eqnarray}
T^{(1)} = \sum_{q} t_q^{(1)} = \sum_{qj} -ie \sqrt{8\pi/3} r_j^{-2} \alpha_j \cdot {\bf Y}_{10
}^{(q)}
\end{eqnarray}
\noindent
where $r_j$ is the radial position of the $j^{th}$ electron, $\alpha_j$ is the Dirac matrix and ${\bf Y}_{10}^{(q)}$ is a vector spherical harmonic.\\
 
\begin{table*}
\begin{ruledtabular}
\begin{tabular}{lccccccccc}
 & \textbf{s$_{1/2}$} & \textbf{p$_{1/2}$} & \textbf{p$_{3/2}$} & \textbf{d$_{3/2}$} & \textbf{d$_{5/2}$} & \textbf{f$_{5/2}$} & \textbf{f$_{7/2}$} & \textbf{g$_{7/2}$} & \textbf{g$_{9/2}$} \\
 &  &  &  &  &  &  &  &  &  \\
\hline
 &  &  &  &  &  &  &  &  &  \\
 & 38 & 35 & 35 & 30 & 30 & 25 & 25 & 20 & 20 \\
Active holes & 6 & 4 & 4 & 3 & 3 & 1 & 1 & 0 & 0 \\
Active particles & 7  & 9 & 9 & 8 & 9 & 7 & 7 & 7 & 7 \\
%Bound particles & 3  & 4 & 3 & 3 & 3 & 3 & 3 & 2 & 2 \\
Upper energy limit (a.u.) & 2800 & 2850 & 2850 & 510 & 510 & 22.6 & 22.6  &  22.6 &  22.6 \\
 &  &  &  &  &  &  &  &  &  \\
\hline
\end{tabular}
\end{ruledtabular}
\caption{Description of total number of basis functions, active holes and
  active particles  involved in this calculation}
\label{tab:front}
\end{table*}

\noindent
We have used the RCC theory in to obtain the atomic wavefunctions. As pointed
out in our earlier work \cite{sahoo03} coupled-cluster theory is equivalent to
all order many-body perturbation theory (MBPT).
In the open-shell coupled-cluster theory \cite{lindgren,mukherjee}  the
many-body  wavefunction for a system with single valence electron can be written as
\begin{eqnarray}
|\Psi_v\rangle = e^T \{1+S_v\} a_v^{\dagger} |\Phi_0\rangle\;,
\end{eqnarray}
where $a_v^{\dagger}$ is the creation operator corresponding to a valence
orbital '{\it v}' and $|\Phi_0\rangle$ is a closed-shell determinantal state
built from occupied Dirac-Fock (DF) orbitals. T- and $S_v$- are the closed and open
shell excitation operators respectively. In this work both T- and $S_v$-
operators are truncated beyond double excitations and triple
excitations are added on the leading order MBPT level.\\

\noindent
Explicitly, the T- operator is defined as
\begin{eqnarray}
T &=& T_1 + T_2 \nonumber \\
&=& \sum_{a,p} a_p^{\dagger} a_a t_a^p + \frac {1}{2} \sum_{ab,pq} a_p^{\dagger} a_q^{\dagger} a_b a_a t_{ab}^{pq}
\end{eqnarray}
\noindent
where $t_a^p$ and $t_{ab}^{pq}$ are the amplitudes of the single and double
excitations from the closed-shell core. Similarly, the open-shell excitation
operator ($S_v$) is defined as
\begin{eqnarray}
S_v &=& S_{1v} + S_{2v} \nonumber \\
&=& \sum_{p\ne v} a_p^{\dagger} a_v s_v^p + \frac {1}{2} \sum_{a,pq} a_p^{\dagger} a_q^{\dagger} a_a a_v s_{va}^{pq}
\end{eqnarray}
\noindent
with $s_v^p$ and $s_{va}^{pq}$ being the single and double excitation
amplitudes  involving the valence electron. \\

\noindent
In coupled-cluster theory the expectation value of any operator can be
expressed as \cite{sahoo02}
\begin{eqnarray} \label{eq:expectation}
< O > &=& \frac {<\Psi_v | O | \Psi_v >} {<\Psi_v|\Psi_v>} \nonumber \\
 &=& \frac {< \Phi_v | \{1+S_v^{\dagger}\} e^{T^{\dagger}} O e^T \{1 + S_v\} | \Phi_v>} { < \Phi_v | \{1+ S_v^{\dagger}\} e^{T^{\dagger}} e^T \{1+ S_v\} | \Phi_v> }
\end{eqnarray}
The above expression was applied to compute the hyperfine structure constant 
{\it 'A'} as given in eqn.\ (\ref{eq:hfsc}).\\ 

\noindent
The orbitals used in the present work were constructed as linear combinations
of Gaussian type orbitals(GTOs) of the form \cite{rajat02}
\begin{equation}
F_{i,k}(r) = r^k . e^{-\alpha_ir^2}
\end{equation}
where $k = 0,1,...$ for s,p,.. type orbital symmetries respectively. For the exponents, the even tempering condition 
\begin{equation}
\alpha_i = \alpha_0 \beta^{i-1}
\end{equation}
was used. The occupied orbitals are the DF single particle states for
closed-shell $Pb^{++}$. The virtual $V^{N-1}$ orbitals \cite{kelly,perger} were
constructed from the closed-shell potential of $Pb^{++}$ using the same
Fock operator. All orbitals
were generated on a grid using a two-parameter Fermi nuclear distribution
approximation given  by
\begin{equation}
\rho = \frac {\rho_0} {1 + e^{(r-c)/a}}
\end{equation}

\begin{figure}
\label{fig:goldstone}
\includegraphics[width=8.0cm]{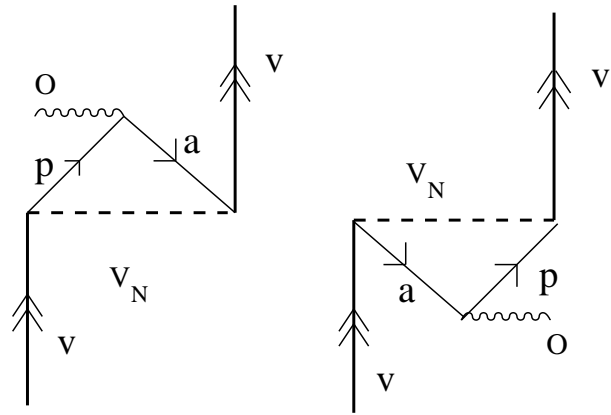}
\caption{Goldstone diagrams representing core-polarization in MBPT}
\end{figure}

\noindent
where the parameter 'c' is the {\it half-charge radius,} and 'a' is related to
the {\it skin thickness} which is defined as the interval of the nuclear
thickness which the nuclear charge density falls from near one to near  zero
The number of basis functions used for generating the occupied and virtual
orbitals are given in table I. The active virtual ('particle') and 
core ('hole') electrons
considered for the coupled-cluster calculations are also displayed. The upper
energy limits above which the virtual orbitals  were truncated during the
RCC computation are given in atomic units. We have chosen $\alpha_0$ as 0.00825 and
$\beta$ as 2.73 for all the symmetries for the generation of the GTO's.\\

\begin{table}
%\begin{ruledtabular}
\begin{tabular}{lcc}
\hline
\hline
\textbf{States} & \textbf{DF approximation}  & \textbf{Experiment} \cite{werth01}\\
\hline
 & & \\
$6p_{1/2}$ & 11513.5 & 13000 \\
$6p_{3/2}$ & 918.3  & 583(21) \\
$7s_{1/2}$ & 7822.9 & -  \\
$7p_{1/2}$ & 1983.1  & - \\
 & & \\
\hline
\hline
\end{tabular}
%\end{ruledtabular}
\caption{Dirac-Fock and Experimental results for magnetic dipole hyperfine
  structure constants  of Pb$^+$ in MHz}
\label{tab:front}
\end{table}

\begin{table}
%\begin{ruledtabular}
\begin{tabular}{lcccc}
\hline
\hline
\textbf{Lowest order} & \textbf{6p$_{1/2}$} & \textbf{7p$_{1/2}$} & \textbf{6p$_{3/2}$} & \textbf{7s$_{1/2}$} \\
\textbf{MBPT terms} &  \textbf{state} & \textbf{state} & \textbf{state} & \textbf{state} \\
\hline
 & & & & \\
 O(DF) & 11513.5 & 1983.1 & 918.3 & 7822.9 \\
RMBPT(2) & 15722.8 & 2578.4 & 302.9 & 12663.9 \\
 core-pol. & 1506.2 & 82.1 & -814.6 & 1624.21 \\
 pair-corr. & 2297.4 & 359.6 & 203.6 & 3012.7 \\
 & & & & \\
\hline
\hline
\end{tabular}
%\end{ruledtabular}
\caption{Second order MBPT results for Pb$^+$ hyperfine structure constants in
  MHz (second row) and the dominating contributions (third and fourth row)
as shown in fig.1\ \protect\ref{fig:goldstone}.}
\label{tab:front1}
\end{table}

\begin{table}
%\begin{ruledtabular}
\begin{tabular}{cccc}
\hline
\hline
\textbf{Virtual orbital} & \textbf{Core orbital}  & \textbf{RMBPT(2)} & \textbf{RCCT}\\
\hline
 & & \\
$7s_{1/2}$ & $6s_{1/2}$ & -190.88 & -128.28\\
$8s_{1/2}$ & $6s_{1/2}$ & -51.16  & -35.99\\
$9s_{1/2}$ & $6s_{1/2}$ & -169.96 & -123.44\\
$10s_{1/2}$ & $6s_{1/2}$ & -468.88& -369.62\\
$11s_{1/2}$ & $6s_{1/2}$ & -90.62 & -73.46\\
 & & \\
\hline
\hline
\end{tabular}
%\end{ruledtabular}
\caption{Contributions of the $6s_{1/2}$ core electron (in MHz) to the
  core-polarization effect using the RMBPT(2) approximation 
(third column) and RCC theory (forth column)}
\label{tab:front2}
\end{table}

\begin{table}
%\begin{ruledtabular}
\begin{tabular}{lcccc}
\hline
\hline
\textbf{Terms} & \textbf{6p$_{1/2}$} & \textbf{6p$_{3/2}$} & \textbf{7s$_{1/2}$} & \textbf{7p$_{1/2}$} \\
 & \textbf{state} & \textbf{state} & \textbf{state} & \textbf{state} \\
\hline
 & & & & \\
 O & 11513.5 & 918.3 & 7822.9 & 1983.1\\
%$\bar O$ & 10848.2 & 962.0 & 6839.6 & 1897.7\\
$O - \bar O$ & 665.3 & -43.7 & 983.3 & 85.4 \\
$\bar O S_{1v} + cc $ & 952.2 & 78.4 & 2122.6 & 326.6\\
$\bar O S_{2v} + cc $ & 1188.2 & -591.0 & 1916.8 & 35.6\\
$S_{1v}^{\dagger} \bar O S_{1v}$ & 21.0  & 1.6 & 164.6 & 14.1 \\
$S_{1v}^{\dagger} \bar O S_{2v} + cc $ & 22.2 & 0.6  & 180.2& 19.2 \\
$S_{2v}^{\dagger} \bar O S_{2v} + cc $ & 149.6 & 194.61 & 298.8 & 18.7\\
\hline\\
\multicolumn{5}{c}{\textbf{Important effective two-body terms of $\bar O$ }} \\
\hline
 & & & & \\
$S_{2v}^{\dagger} O T_1 + cc $ & -20.2 & 2.0 & 14.6 & -0.76\\
$S_{2v}^{\dagger} O T_2 + cc $ & -160.2 & -12.6 & -135.4 & -21.64\\
 Norm. & -88.5 & -6.7 & -181.8 & -22.98\\
\hline\\
 Total & 12903.7 & 623.2 & 11158.6 & 2263.5\\
Experiment & 13000 & 583(21) &     &       \\
\hline
\hline
\end{tabular}
%\end{ruledtabular}
\caption{Contribtions of different coupled-cluster terms to the Pb$^+$
  hyperfine stucture constant.  $cc$ stands for the complex conjugate part of the corresponding terms}
\label{tab:front3}
\end{table}

\begin{figure}
\includegraphics[width=8.0cm]{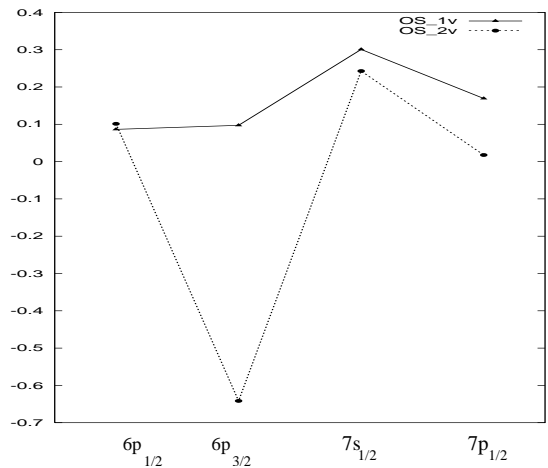}
\caption{The ratios of pair-correlation and core-polarization effects
 w.r.t. the DF values}
\end{figure}

\noindent
In table II we present the results for the hyperfine constants
using the DF approximation
and compare them with the experimental values for the
$6p_{1/2}$ and $6p_{3/2}$ states -- the only two states on which measurements
have been made. The poor agreement of the results indicate
the importance of correlation effects for these states (which
were absent in the DF approximation). It
is interesting to note that the DF values for these two states deviate from
their respective experimental values in opposite directions,
so that the sign of the correlation contributions are opposite for the
two cases. This is further supported by the results based on second order
relativistic many-body perturbation theory (RMBPT(2)) which are
given in table III. Here the dominant contributions to RMBPT(2)
as shown in Fig.\ \ref{fig:goldstone} are given explicitly. 
While electron correlation at this level is substantial
for all the states, it is dramatic in the case of $6p_{3/2}$ state because of
the unusually large and negative core-polarization. In
table IV we give the significant contributions to the core polarization which arises
from the interaction of the outermost core 6s and the valence $6p_{3/2}$
electrons (third column). The sum of these individual contributions is -971.5 MHz and after
taking into account the polarization of all the other core electrons, a net
contribution of -814.6 MHz is obtained. The tremendous size of this 
second-order correction suggests that an all order method
like coupled-cluster theory is necessary for a correct quantitative
description of the correlation effects in $Pb^+$. 
This is indeed reflected in the results given in table  V.
Again, the 'bare' operator $O$ represents the DF approximation,
i.e.\ excluding any correlation effects. Several important
correlation contributions for our RCC calculations are also presented
in table V.  $\bar O =  e^{T^{\dagger}} O e^T$ denotes the 'dressed'
operator containing the contributions of the closed-shell 
cluster amplitudes in Eq.\ (\ref{eq:expectation}). 
Although core-polarization ($OS_2v$) and pair correlation ($OS_1v$) are the dominant
correlation effects, core correlation effects ($\bar O - O$) are by
no means negligible; they amount to about 9\% for the 7s  state.
Summing up all the contributions given in table V leads to
significant improvements in our calculated values of the hyperfine constants
of the $6p_{1/2}$ and $6p_{3/2}$ states. The sub one percent (0.7\%) agreement between the former and experiment is indeed spectacular. A similar agreement cannot be expected for the
latter state which is characterised by extremely peculiar correlation
effects. Even so, the hyperfine constant for this state differs from
experiment (3.6\% error bar) by a little less than 7\%. This is certainly
remarkable considering that the corresponding discrepancy at the level of
RMBPT(2) is 48\%. It is interesting to note from table IV 
(fourth column) that the core-polarization contributions in RCC 
theory follow the same trend as in  RMBPT(2).\\

\noindent
The plot in fig.\ 2 highlights the relative importance of the core-polarization
and pair correlation for the different states. It is instructive to point out
that, unlike the hyperfine constant in the ground state of
$Ba^+$ \cite{sahoo02}, core-polarization effects are larger than
pair-correlation for the ground and first excited state, i.e. $6p_{1/2}$ and
$6p_{3/2}$ states of $Pb^+$. This is the result of the much stronger
valence-core interactions in $Pb^+$ compared to  $Ba^+$.\\

\noindent
In summary, the strength of RCC theory has been exploited to obtain
for the first time a quantitative understanding of the interplay of
relativistic and correlation effects in the magnetic dipole hyperfine
constants for Pb$^+$. It has been demonstrated that the results of
the DF and RMBPT(2) approximations differ substantially from the
measured values of the hyperfine constants. However, the inclusion
of single, double and a subset of triple excitations to all orders
in the framework of RCC theory leads to a dramatic improvement in
the results. The relevance of the present work extends beyond 
hyperfine interactions in Pb$^+$. Our results highlight the fact that
a judicious use of RCC theory can yield accurate results for 
properties that are sensitive to the nuclear region. Indeed, this
has important implications for Tl, which like Pb$^+$ is a heavy atomic
system with strongly interacing configurations and is one of the
leading candidates for the study of parity nonconservation due to
neutral weak currents \cite{merlitz,vetter95,dzuba87}.\\ 

\noindent
We are grateful to Prof. Werth for his valuable discussions and suggestion for
this calculation. It was possible to contact with Prof. Werth through DST-DAAD
exchange programme. The calculation is carried out using the Tera-flopp
Supercomputer in C-DAC,  Bangalore.


\begin{thebibliography}{14}
\bibitem{bishop}
F. Bishop Raymond, {\it Microscopic Quantum Many-body Theories and their Applications}, Edited by J. Navarro and A. Polls, page. 1, Springer-Verlag Berlin Heidelberg (1998)
\bibitem{kaldor01}
U. Kaldor, {\it Microscopic Quantum Many-body Theories and their Applications}, Edited by J. Navarro and A. Polls, page. 71 (1998)
\bibitem{kaldor02}
U. Kaldor, {\it Recent Advances in Coupled-cluster Methods}, Edited by R. J. Bartlett, page. 125, World Scientific, Singapore (1997)
\bibitem{merlitz}
H. Merlitz, G. Gopakumar, R. K. Chaudhuri, B. P. Das, U. S. Mahapatra and D. Mukhrjee, Phys. Rev. A {\bf 63}, 025507 (2001)
\bibitem{strumia}
F. Strumia, Proceedings of the 32nd. Annual Symposium on Frequency Control, Atlantic City, N.J., USA 1978
\bibitem{werth01}
A. Roth, G. Werth, Z. Phys. D- Atoms, Molecules and Clusters {\bf 9}, 265(1988)
\bibitem{werth02}
Xin Feng, Guo-Zhong Li, R. Alheit, and G. Werth, Phys. Rev.A {\bf 46}, 327(1992)
\bibitem{sahoo01}
B.K. Sahoo, R. K. Chaudhuri, B.P.Das, S. Majumder, H. Merlitz, U.S. Mahapatra and D. Mukherjee, J. Phys. B {\bf 36}, 1899(2003)
\bibitem{sahoo02}
B.K. Sahoo, G. Gopakumar, R. K. Chaudhuri, B.P.Das, H. Merlitz, U.S. Mahapatra and D. Mukherjee, Phys. Rev. A {\bf 68}, 040501(R) (2003)
\bibitem{csur01}
Chiranjib Sur, B. K. Sahoo, Rajat K. Chaudhuri, B. P. Das and D. Mukherjee, (submitted to Euro. J. Phys. D) arXiv:physics/0310098 (2003)
\bibitem{cheng}
K.T. Cheng and W.J.Childs, Phys. Rev. A {\bf 31}, 2775 (1985)
\bibitem{sahoo03}
B. K. Sahoo, S. Majumder, R. K. Chaudhuri, B. P. Das, D. Mukherjee, (in press(J. Phys. B)), arXiv:physics/0403134 2004)
\bibitem{lindgren}
I. Lindgen and J. Morrison, {\it Atomic Many-Body Theory}, edited by G. Ecker, P. Lambropoulos, and H. Walther ( Springer-Verlag, Berlin, 1985)
\bibitem{mukherjee}
D. Mukherjee and S. Pal, Adv. Quantum Chem. {\bf 20}, 281 (1989)
\bibitem{geetha02}
 Geetha Gopakumar, Holger Merlitz,Rajat K. Chaudhuri, B. P. Das,Uttam Sinha Mahapatra and Debashis Mukherjee, Phys. Rev. A {\bf 66}, 032505 (2002)
\bibitem{rajat02}
R. K. Chaudhuri, P. K. Panda, H. Merlitz, B. P. Das, U. S. Mahapatra and D Mukhe
rjee, {\it J. Phys. B}, {\bf 33}, 5129 (2000)
\bibitem{kelly}
H. P. Kelly, Phys. Rev. {\bf 131}, 684 (1963)
\bibitem{perger}
W. F. Perger and B. P. Das, Phys. Rev. A {\bf 35}, 3942 (1987)
\bibitem{vetter95}
P. A. Vetter, D. M. Meekhof, P. K. Majumder, S. K. Lamoreaux and
E. N. Fortson, Phys. Rev. Lett. {\bf 74}, 2658 (1995)
\bibitem{dzuba87}
V. A. Dzuba, V. V. Flambaum, P. G. Silverstrov and O. P. Sushkov, 
J. Phys. B {\bf 20}, 3297 (1987)
\end{thebibliography}
\end{document}